\documentclass[aps,prd,twocolumn,amsmath,amssymb,nofootinbib,floatfix]{revtex4-2}

\usepackage{graphicx}
\usepackage{amsmath}
\usepackage{amssymb}
\usepackage{bm}
\usepackage{hyperref}

\begin{document}

\title{Higher-order local constraints from reciprocal symmetry \\ and entanglement entropy of charged-particle multiplicity distributions in $pp$ collisions}

\author{Mustapha Ouchen}
\email{mustapha.ouchen@msmail.ariel.ac.il}
\affiliation{Department of Physics, Ariel University, Ariel 4077601, Israel}

\author{Alex Prygarin}
\email{alexanderp@ariel.ac.il}
\affiliation{Department of Physics, Ariel University, Ariel 4077601, Israel}

\author{Claudelle Capasia Madjuogang Sandeu}
\email{claudell.madjuoga@msmail.ariel.ac.il}
\affiliation{Department of Physics, Ariel University, Ariel 4077601, Israel}

\begin{abstract}
The KNO-violating term $f_s$ of the charged-particle multiplicity distribution in $pp$ collisions measures the relative deviation of $\langle n\rangle P_n$ from $e^{-z}$, with $z=n/\langle n\rangle$, and is reported to obey the reciprocal symmetry $f_s(z)=f_s(1/z)$, taken here as input. Being an evenness condition in $\ln z$, it generates a tower of local constraints on the derivatives of $P_n$ at the mean. The lowest member holds for the ATLAS data at $7$, $8$ and $13$~TeV at the few-per-cent level, while the third-derivative residual testing the next member is not determined with a controlled uncertainty by the present binning and stays inconclusive. The global test is consistent at $7$ and $8$~TeV, while at $13$~TeV a residual deviation remains, which a closure test shows is not a binning artefact. Multiplicative noise in the Mueller colour-dipole cascade, the negative binomial and the dipole cascades with recombination each give an $f_s$ that is not invariant under $z\to1/z$, so none of them carries the symmetry. We further obtain a dynamics-independent entropy in the KNO continuum, $S=\ln\langle n\rangle+I_0-\tfrac12\int e^{-z}f_s^2\,dz$, with $I_0$ a support factor tending to unity in the continuum limit. The linear term cancels by normalisation and unit mean, independently of the symmetry, and the remaining correction is quadratic and negative. It reproduces the Shannon entropy of the ATLAS data at the $10^{-3}$ level.
\end{abstract}

\maketitle

\section{Introduction}

Charged-particle multiplicity distributions $P_n$ in high-energy hadron collisions enjoy KNO scaling~\cite{KNO:1972}: at large mean multiplicity the combination $\langle n\rangle P_n$ depends on the energy only through the scaled variable $z=n/\langle n\rangle$ and approaches a universal KNO function $\psi(z)$. The cascade of the Mueller colour-dipole model~\cite{Mueller:1994,Mueller:1995} gives a simple prediction for this function, the geometric law $\psi(z)=e^{-z}$, which the data follow only approximately. The deviation from the geometric baseline is measured by the KNO-violating term $f_s$, defined as the relative deviation of $\langle n\rangle P_n$ from $e^{-z}$ (so that $f_s=0$ corresponds to the exact cascade result); the precise definition is given in Sec.~\ref{sec:tower}.

It was reported in~\cite{Ouchen:2026} that $f_s$ is not an arbitrary function of $z$, but appears to respect the reciprocal symmetry $f_s(z)=f_s(1/z)$. The symmetry is observed in the ATLAS charged-multiplicity data at $\sqrt{s}=7$, $8$ and $13$~TeV and in the CMS data at $7$~TeV~\cite{Ouchen:2026,CMS-7TeV}, over the window $1/3<z<3$, where $f_s$ is well described by a Gaussian in $\ln z$ centred at $z=1$. We take this symmetry as input in the present paper. The lowest consequence of the symmetry extracted in~\cite{Ouchen:2026} is the local relation $P'(\langle n\rangle)=-P(\langle n\rangle)/\langle n\rangle$ at $n=\langle n\rangle$. The scaling point $z=2$, where KNO scaling is approximately restored across these energies, plays a special role and was analysed separately in~\cite{Ouchen:2025MEM}.

The Shannon entropy of the multiplicity distribution is of interest in this context for a separate reason. In the proposal of Kharzeev and Levin~\cite{Kharzeev:2017} it is identified, at small $x$, with the entanglement entropy of the partonic state probed in deep inelastic scattering, an identification that has been compared with the H1 data~\cite{H1:2021} and developed for BFKL-evolved gluon distributions~\cite{Hentschinski:2022}. For the maximally entangled state this entropy is fixed by the mean multiplicity, $S\simeq\ln\langle n\rangle$. It is natural to ask how the KNO-violating term $f_s$ enters this number once the distribution departs from the exact exponential.

In the present paper we take the reciprocal symmetry of Ref.~\cite{Ouchen:2026} as the starting point. The symmetry $f_s(z)=f_s(1/z)$ is the statement that $h(u)\equiv f_s(e^u)$ is even in $u=\ln z$, so that all odd derivatives of $h$ at $u=0$ vanish. This generates a tower of local constraints on the derivatives of $P_n$ at $n=\langle n\rangle$, of which the relation of~\cite{Ouchen:2026} is the lowest ($k=0$) member. We work out the next ($k=1$) member and the third-derivative residual it controls, and we test the tower in the ATLAS data. We then ask whether the symmetry can be generated by the cascade dynamics itself, and find that multiplicative noise on the branching rate of the Mueller colour-dipole model gives $f_s\sim z^2-4z+2$, which is not invariant under $z\to1/z$; the negative-binomial distribution and two-component geometric mixtures share this feature. Finally we obtain a dynamics-independent expression (within the KNO continuum framework) for the entanglement entropy in terms of $f_s$, in which the term linear in $f_s$ cancels by normalisation and $\langle z\rangle=1$, leaving a small negative-definite deficit. This formula, which casts the entropy of the multiplicity distribution in the same KNO-violation language as the symmetry itself, is one of the main results of the present paper.

\section{Overview of the main results}
\label{sec:overview}

The reciprocal symmetry $f_s(z)=f_s(1/z)$ of the KNO-violating term, with $f_s$ defined in \eqref{eq:fs} and $z=n/\langle n\rangle$, was found in the ATLAS and CMS multiplicity data over the limited window $1/3<z<3$ (or narrower)~\cite{Ouchen:2026}. The present paper does not establish the symmetry but takes it as input and works out three consequences. The symmetry is the statement that $h(u)\equiv f_s(e^u)$ is even in $u=\ln z$, so that every odd derivative $h^{(2k+1)}(0)$ vanishes, and each vanishing is one local algebraic constraint on the derivatives of $P_n$ at $n=\langle n\rangle$. This generates a tower, of which the relation $P'(\langle n\rangle)=-P(\langle n\rangle)/\langle n\rangle$ of~\cite{Ouchen:2026} is the lowest ($k=0$) member. That member is a weak indicator (it is the condition $f_s'(1)=0$, satisfied by any $f_s$ stationary at $z=1$, equivalently by any KNO shape whose $e^{z}\langle n\rangle P_n$ is extremal at the mean), and the new content of the present paper is the next ($k=1$) member,
\begin{equation*}
\langle n\rangle^{3}P'''(\langle n\rangle)+6\,\langle n\rangle^{2}P''(\langle n\rangle)=5\,P(\langle n\rangle),
\end{equation*}
quoted from \eqref{eq:k1}, which is the reduced form of the $k=1$ condition once the $k=0$ relation is imposed. The unconditional test is the third-derivative residual $\delta_3=1-2\rho_0+\rho_1$ of \eqref{eq:delta3}, which vanishes if and only if $h^{(3)}(0)=0$, whatever the value of $\rho_0$. We test the tower on the ATLAS data: $\rho_0$ sits close to unity at all three energies ($\rho_0=0.975\pm0.006$ at $13$~TeV), while $\delta_3$ cannot be extracted with a controlled uncertainty (it crosses zero monotonically with the fit window), so the $k=1$ test is left inconclusive by the present binning.

We then ask whether the cascade dynamics can generate the symmetry by itself. Multiplicative noise on the branching rate of the Mueller colour-dipole model gives the parameter-free shape $f_s(z)\propto z^2-4z+2$ of \eqref{eq:noise-fs}, which respects the two normalisation conditions but is not invariant under $z\to1/z$ (its roots $z=2\pm\sqrt2$ multiply to $2$ rather than forming reciprocal pairs). The negative-binomial distribution and two-component geometric mixtures share this feature. The reciprocal symmetry is therefore not generated by simple rate fluctuations, by the negative binomial, or by two-component mixtures, and the recombination and conformal-weight cascades of~\cite{Kutak:2025,Kutak:2025-2} reduce to a negative binomial ($\delta_3=(1-k)/5$) that does not carry it either. The tower thus filters candidate mechanisms: none of these schemes carries the symmetry, and its dynamical origin is to be sought beyond them.

The third result is a dynamics-independent expression (within the KNO continuum framework), in terms of $f_s$, for the Shannon entropy of the multiplicity distribution, which Kharzeev and Levin~\cite{Kharzeev:2017} identify with the entanglement entropy of the partonic state,
\begin{equation*}
S=\ln\langle n\rangle+I_0-\tfrac12\!\int_{\mathcal S}e^{-z}f_s^2(z)\,dz+\mathcal O(f_s^3),
\end{equation*}
with $I_0\equiv\int_{\mathcal S}e^{-z}dz$, which is \eqref{eq:S-formula}. The term linear in $f_s$ cancels by normalisation and $\langle z\rangle=1$, and the remaining quadratic piece is a negative-definite deficit $\Delta S\le0$. The cancellation rests only on normalisation and $\langle z\rangle=1$ and does not use the reciprocal symmetry, so this result is independent of, and complementary to, the tower. It accounts for the empirical $S\simeq\ln\langle n\rangle+1$ without assuming an exponential distribution (the $+1$ is the entropy $I_0$ of the geometric KNO shape $e^{-z}$~\cite{Mueller:1994,Mueller:1995} in the continuum limit, not the maximally entangled value $\ln\langle n\rangle$), and on the ATLAS data it reproduces the direct Shannon entropy at the $10^{-3}$ level with a small deficit $|\Delta S|\lesssim0.015$.

\section{Tower of local constraints}
\label{sec:tower}

Charged-particle multiplicity distributions $P_n$ at high energy obey KNO
scaling~\cite{KNO:1972,DreminGary:2001}: at large $\langle n\rangle$ they depend
on $n$ only through $z=n/\langle n\rangle$,
\begin{equation}
\langle n\rangle P_n=\psi(z),\qquad z=n/\langle n\rangle,
\end{equation}
with $\int_0^\infty\psi\,dz=1$ and $\int_0^\infty z\,\psi\,dz=1$.

The cascade of the Mueller colour-dipole model~\cite{Mueller:1994,Mueller:1995} gives $\psi(z)=e^{-z}$ at
leading order in $1/\langle n\rangle$, with subleading $1/\langle n\rangle$
corrections~\cite{Ouchen:2025AGK}. The deviation from this leading exponential is
measured by~\cite{Ouchen:2026}
\begin{equation}
f_s(z)\equiv\frac{\langle n\rangle P_n-e^{-z}}{e^{-z}}=\psi(z)\,e^{z}-1,
\label{eq:fs}
\end{equation}
so that $\psi(z)=e^{-z}[1+f_s(z)]$. The two normalisation conditions become
\begin{equation}
\int_0^\infty e^{-z}f_s(z)\,dz=0,\qquad\int_0^\infty z\,e^{-z}f_s(z)\,dz=0 .
\label{eq:fs-norm}
\end{equation}
In the ATLAS data (and the CMS data at $7$~TeV~\cite{Ouchen:2026,CMS-7TeV}, a non-single-diffractive sample with $|\eta|<2.4$) $f_s$ satisfies the reciprocal symmetry
$f_s(z)=f_s(1/z)$ over $1/3<z<3$~\cite{Ouchen:2026}; the point $z=2$ is special,
where KNO scaling is approximately restored across energies~\cite{Ouchen:2025MEM}.
A complete analysis of $f_s$ and its comparison with the experimental data was
carried out with high accuracy in~\cite{Ouchen:2026,Ouchen:2025MEM}.

The lowest local consequence of the symmetry, a relation between $P_n$ and its
first derivative at $n=\langle n\rangle$, was found in~\cite{Ouchen:2026},
\begin{equation}
P'(\langle n\rangle)=-\frac{P(\langle n\rangle)}{\langle n\rangle},
\label{eq:k0prev}
\end{equation}
and is in very good agreement with the experimental data~\cite{Ouchen:2026}.

In this paper we exploit the reciprocal symmetry and the other properties of
$f_s$ to generalise this relation to higher derivatives of $P_n$ at
$n=\langle n\rangle$, and to derive a dynamics-independent expansion of the entropy (within the KNO continuum) in terms of $f_s$.

The reciprocal symmetry $f_s(z)=f_s(1/z)$ is equivalent to the statement that
$h(u)\equiv f_s(e^u)$ is an even function of $u=\ln z$, so that all its odd-order
derivatives vanish at $u=0$. This generates a tower of local constraints on the
higher derivatives of $P_n$ at $n=\langle n\rangle$, of which \eqref{eq:k0prev} is
the lowest ($k=0$) member. The derivation is given in Appendix~\ref{app:tower}.
The next ($k=1$) member is
\begin{equation}
\langle n\rangle^{3}P'''(\langle n\rangle)+6\,\langle n\rangle^{2}P''(\langle n\rangle)\;=\;5\,P(\langle n\rangle),
\label{eq:k1}
\end{equation}
a relation among the value, second, and third derivatives of $P_n$ at $n=\langle n\rangle$. Since $P_n$ is discrete and $\langle n\rangle$ is non-integer, these derivatives are those of a smooth KNO interpolation of the measured distribution near the mean (Appendix~\ref{app:fit}), so that the tower constrains any such smooth interpolation that realises the reciprocal symmetry.

\subsection{Dimensionless ratios}

For experimental tests, it is convenient to define dimensionless ratios that equal unity if the symmetry holds. The $k=0$ ratio is~\cite{Ouchen:2026}
\begin{equation}
\rho_0\;\equiv\;-\langle n\rangle\,\frac{P'(\langle n\rangle)}{P(\langle n\rangle)},\qquad \rho_0\stackrel{\rm sym.}{=}1.
\label{eq:rho0}
\end{equation}
The $k=1$ counterpart, derived from \eqref{eq:k1}, is
\begin{equation}
\rho_1\;\equiv\;\frac{1}{5}\!\left[\frac{\langle n\rangle^{3}P'''(\langle n\rangle)}{P(\langle n\rangle)}+6\,\frac{\langle n\rangle^{2}P''(\langle n\rangle)}{P(\langle n\rangle)}\right],\qquad \rho_1\stackrel{\rm sym.}{=}1.
\label{eq:rho1}
\end{equation}
A test of the symmetry beyond the leading derivative reduces to an experimental determination of $\rho_1$ from the multiplicity distribution near $n=\langle n\rangle$. We also consider the third-derivative residual derived in Appendix~\ref{app:tower},
\begin{equation}
\delta_3 \;\equiv\; 1-2\rho_0+\rho_1\;\stackrel{\rm sym.}{=}\;0,
\label{eq:delta3}
\end{equation}
which vanishes if and only if $h^{(3)}(0)=0$ holds in the data, irrespective of how exactly the $k=0$ relation is satisfied. The independent test of the higher-order constraint is $\delta_3=0$.

\begin{figure*}[t]
\includegraphics[width=\textwidth]{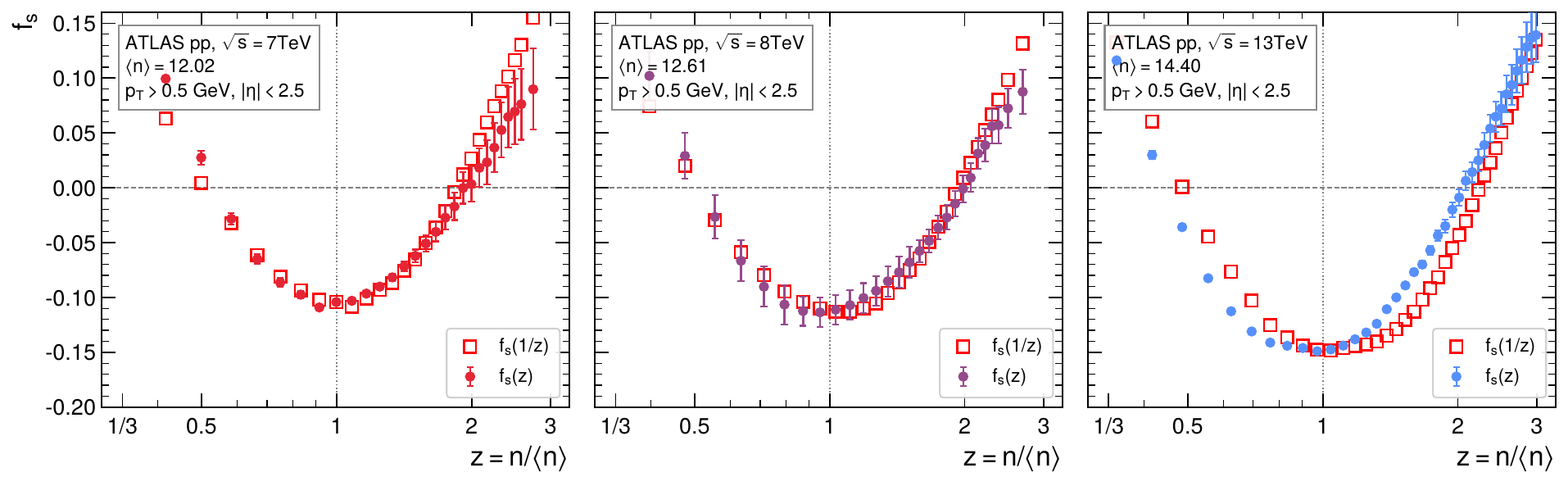}
\caption{The function $f_s(z)$ (filled circles) and $f_s(1/z)$ (open squares) plotted against $z$ in the window $1/3<z<3$, for ATLAS $pp$ data at $\sqrt{s}=7$, $8$ and $13$ TeV. The two curves coincide within experimental uncertainties at $7$ and $8$ TeV; at $13$ TeV the smaller error bars make visible a residual scatter away from $z=1$; the associated $\chi^2_{\rm sym}$ (Sec.~\ref{sec:test}) is large, and a closure test (Appendix~\ref{app:fit}) shows it is not a binning artefact.}
\label{fig:symmetry}
\end{figure*}

Figure~\ref{fig:symmetry} shows $f_s(z)$ and $f_s(1/z)$ in the window $1/3<z<3$ for the three energies. At $7$ and $8$~TeV the two curves coincide within experimental uncertainties, illustrating the reciprocal symmetry that underlies the tower, while at $13$~TeV a residual deviation away from $z=1$ remains visible (Sec.~\ref{sec:test}).

\subsection{Test on the ATLAS data}
\label{sec:test}

We test the constraints on the ATLAS charged-multiplicity
data~\cite{ATLAS-7TeV,ATLAS-8TeV,ATLAS-13TeV} at $\sqrt{s}=7$, $8$ and $13$~TeV (charged particles, $p_T>0.5$~GeV, $|\eta|<2.5$, $n_{\rm ch}\ge1$),
extracting $\rho_0$, $\rho_1$ and $\delta_3$ from local polynomial fits near
$n=\langle n\rangle$ (Appendix~\ref{app:fit}). The $k=0$ ratio $\rho_0$ is
close to unity at all three energies ($\rho_0=0.972$, $0.953$ and $0.975$ at $7$, $8$ and $13$~TeV for $W=6$, with within-fit errors $0.011$, $0.051$ and $0.006$; the window dependence adds a further spread of $\sim0.02$),
confirming~\eqref{eq:k0prev}. The $k=1$ residual $\delta_3$, however, cannot be
extracted with a controlled uncertainty: at $13$~TeV it varies monotonically with
the fit window, $\delta_3=+0.79,+0.31,-0.02,-0.17,-0.30$ for $W=4$--$8$
(Fig.~\ref{fig:robustness} in Appendix~\ref{app:fit}), so the apparent agreement at
$W=6$ is a zero-crossing and not a stable result.

The global test of $f_s(z)=f_s(1/z)$ over $1/3<z<3$ gives
$\chi^2_{\rm sym}/N_{\rm pairs}=1.5,\,0.4,\,27$ at $7$, $8$ and $13$~TeV
($N_{\rm pairs}=20,\,20,\,29$) [Eq.~\eqref{eq:chi2-sym}]. The symmetry is consistent
with the data at $7$ and $8$~TeV. At $13$~TeV, where the errors are smallest, the value
is large, and a closure test (Appendix~\ref{app:fit}) shows that this is not an artefact
of the binning or the interpolation; it reflects a residual pointwise deviation from the
symmetry, consistent with the unstable $\delta_3$ found above. Whether the deviation is a
genuine small breaking at the highest energy or a $z$-dependent correlated systematic,
which ATLAS does not publish differentially, cannot be settled here.

\section{Multiplicative noise on the Markov branching process of the Mueller colour-dipole model}
\label{sec:cascade}

In the pure colour-dipole cascade the KNO function is the geometric law
$\langle n\rangle P_n\to e^{-z}$, for which $f_s\equiv0$. A nonzero $f_s$ therefore
requires a dynamical source of KNO violation. The simplest such source in a
branching process is an event-by-event fluctuation of the cascade rate,
equivalently of the cascade lifetime in rapidity.

The gluon density in a hadron grows rapidly towards small $x$. This growth is
generated by the linear evolution and, if extrapolated far enough, it leads to a
violation of unitarity. The growth is slowed down by non-linear effects, most
notably gluon recombination~\cite{GLR:1983,MuellerQiu:1986}, which restore
unitarity and are resummed, in the large-$N_c$ mean-field limit of the Balitsky
hierarchy, by the Balitsky--Kovchegov
equation~\cite{Balitsky:1996,Kovchegov:1999} (see Ref.~\cite{CGC:review} for a
review). The dilute and the dense regimes are separated by the saturation scale
$Q_s(x)$, at which the gluon occupation number becomes of the order of
$1/\alpha_s$, and this scale grows with energy. In the saturation regime the
produced multiplicity scales with $Q_s^2$, so that an event-by-event shift of
$\ln Q_s^2$ is a shift of $\ln\langle n\rangle$, which is the rate fluctuation
implemented in Appendix~\ref{app:noise}.

The event-by-event diffusion of $\ln Q_s^2$ is itself part of high-energy
evolution. It was studied by Mueller and Shoshi~\cite{Mueller:Shoshi:2004} and in
the wavefront picture of Iancu, Mueller and
Munier~\cite{Iancu:Mueller:Munier:2005}, and it is described by the stochastic
FKPP equation~\cite{Munier:Peschanski:2003,Brunet:Derrida:1997}. We take
multiplicative log-symmetric noise on the cascade rate as the first candidate for
$f_s$ and compute the shape it predicts.

Averaging the geometric cascade~\cite{Mueller:1994,Mueller:1995} over a
Gaussian fluctuation of $\ln Q_s^2$ that rescales the mean gives, to
$\mathcal{O}(\sigma_\eta^2)$ (Appendix~\ref{app:noise}),
\begin{equation}
f_s(z)\big|_{\rm mult.\,noise}=\frac{\sigma_\eta^2}{2}\,(z^2-4z+2),
\label{eq:noise-fs}
\end{equation}
a parameter-free shape whose amplitude is set by the variance of $\ln Q_s^2$. It
satisfies the normalisation conditions $\int e^{-z}f_s\,dz=\int z\,e^{-z}f_s\,dz=0$,
but it is not invariant under $z\to1/z$: its roots $z=2\pm\sqrt2$ are not reciprocal,
and $f_s(2)=-2$ while $f_s(1/2)=1/4$ (in units of $\sigma_\eta^2/2$). Multiplicative
noise on the rate thus gives a KNO-violating term of a shape different from the
symmetric one observed in the data.

\begin{figure}[t]
\includegraphics[width=\columnwidth]{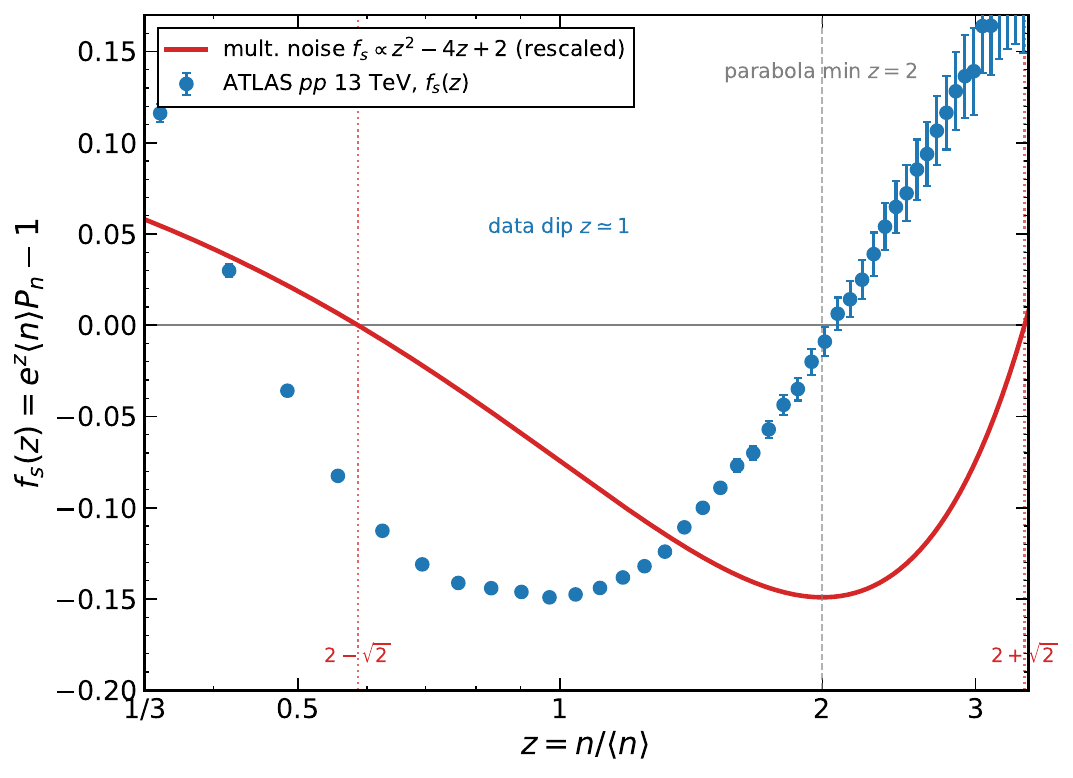}
\caption{Comparison of the multiplicative-noise prediction $f_s\propto z^2-4z+2$ (solid line, rescaled to match the depth of the data dip) with the measured $f_s(z)$ (filled circles, ATLAS $pp$ at $\sqrt{s}=13$ TeV). The polynomial has its minimum at $z=2$ and zeros at $z=2\pm\sqrt2$ (dotted vertical lines), whereas the data dip is at $z\simeq1$; the two shapes do not match.}
\label{fig:cascade}
\end{figure}

Figure~\ref{fig:cascade} compares the multiplicative-noise prediction $f_s\propto z^2-4z+2$ with the ATLAS 13 TeV data. The polynomial has its minimum at $z=2$ and crosses zero at $z=2\pm\sqrt2$, whereas the data dip lies at $z\simeq 1$; once normalised to the same dip depth, the model curve is too broad and its minimum is misplaced, so the shape of the KNO-violating term seen in the data is not reproduced by multiplicative noise on the cascade rate.

Two-component geometric mixtures and the negative binomial behave in the same way.
In the KNO limit neither is invariant under $z\to1/z$ except in trivial cases: the
negative binomial has $\rho_0=1$ for all $k$ but $\delta_3=(1-k)/5$, which is zero
only at the geometric point $k=1$. The AGK extension of the chain of the Mueller colour-dipole model~\cite{Ouchen:2025AGK} has a constant leading $1/\langle n\rangle$ term,
reabsorbed by the rescaling, and an asymmetric subleading $z$-dependence with
intersection points at $z=1/2$ and $z=2$. These examples show how the
tower~\eqref{eq:tower-A} acts as a filter on cascade models: a model that reproduces
the symmetry has to satisfy the relation derived in Appendix~\ref{app:tower} $\xi_3+6\xi_2=5$ at $n=\langle n\rangle$, which points to
correlations beyond independent emission. The recombination and conformal-weight cascades of~\cite{Kutak:2025,Kutak:2025-2} do not carry it. In their KNO limit they reduce to a negative binomial with $\delta_3=(1-k)/5\neq0$, so the symmetry must originate elsewhere (Appendix~\ref{app:noise}).

The inversion $z\to1/z$ resembles the M\"obius (conformal) invariance of the BFKL kernel at
leading logarithmic accuracy~\cite{Lipatov:1986,Lipatov:1993}. The two act on
different spaces: the BFKL $SL(2,\mathbb{C})$ acts on the two-dimensional complex
transverse coordinate of the dipole endpoints, while $z=n/\langle n\rangle$ is a
single positive real variable. The shared name does not imply a shared group action, and absent such an action the analogy is at present purely nominal; a connection between them remains to be established.

\section{Entanglement entropy}
\label{sec:entropy}

The entanglement-entropy programme initiated by Kharzeev and Levin~\cite{Kharzeev:2017} identifies the entropy of the partonic state probed in deep inelastic scattering with the Shannon entropy of the produced-particle multiplicity distribution. For the maximally entangled low-$x$ state the two coincide and reduce to a single number, $S=\ln\langle n\rangle$, fixed by the mean gluon (or hadron) multiplicity. This relation has been compared with HERA data~\cite{Kharzeev:2021,H1:2021}, reviewed in~\cite{Kharzeev:2022}, examined at subnucleonic scales~\cite{Tu:2020}, and developed by Hentschinski, Kutak and collaborators for BFKL-evolved gluon distributions~\cite{Hentschinski:2022}, for charged hadrons~\cite{Hentschinski:Straka:2022}, for diffractive DIS~\cite{Hentschinski:2023}, in a dedicated study of the QCD evolution of the entropy~\cite{Hentschinski:2024}, and against Monte Carlo generators including soft-gluon effects~\cite{Hentschinski:2026}. Dipole-cascade calculations with recombination and transitions to the vacuum~\cite{Kutak:2025,Kutak:2025-2,Kutak:2025-3,Kutak:2026} work in the same setting; a diffusion-scaling description in which the high-multiplicity tails depart from KNO scaling has also been developed~\cite{Moriggi:2025}. A recurrent finding across these analyses is that the data are described by $S\simeq\ln\langle n\rangle$, or by $S\simeq\ln\langle n\rangle+1$ once the geometric (exponential) shape of the cascade distribution is folded in. The additive constant is the entropy of the normalised exponential $\psi(z)=e^{-z}$ that the cascade of the Mueller colour-dipole model produces at large $N_c$~\cite{Mueller:1994,Mueller:1995,DKMT:1991}.

The measured KNO-scaled distribution $\langle n\rangle P_n=\psi(z)$ is not exactly exponential. The deviation $f_s(z)=\psi(z)e^{z}-1$ of Eq.~\eqref{eq:fs} reaches $|f_s|\sim0.2$--$0.4$ in the lowest-multiplicity bins ($z<1/3$), and up to $\sim10$--$15\%$ within the central window $1/3<z<3$. One would expect a correction of comparable size in the entropy. It does not appear, and below we turn this into a dynamics-independent (within the KNO continuum), data-driven measure of the entropy deficit relative to the geometric-cascade value.

We start from the Shannon entropy of the distribution,
\begin{equation}
S=-\sum_n P_n \ln P_n,
\end{equation}
which Kharzeev and Levin identify with the von~Neumann entanglement entropy of the partonic state~\cite{Kharzeev:2017} under two conditions: the reduced density matrix is diagonal in the relevant Fock basis at leading order, and the hadronic multiplicity tracks the partonic one. Both are leading-order statements (the density matrix acquires off-diagonal interference at next-to-leading order), and the identification is further loosened by the choice of charged versus total multiplicity and by the pseudorapidity window~\cite{Hentschinski:Straka:2022,Hentschinski:2024}. We adopt this terminology for the Shannon entropy throughout. A complementary line probes spin-spin Bell-type entanglement of quark--antiquark pairs, in diffractive heavy-quark production~\cite{Fucilla:2025kit}, in exclusive DIS via Generalized Parton Distributions~\cite{Hatta:Schoenleber:2025}, and in inclusive heavy-quark electroproduction~\cite{Fucilla:Hatta:Xiao:2026}; these probe distinct facets of the entanglement structure of small-$x$ QCD and are not addressed here.

In the continuum approximation $\sum_n\to\int dn$, with $P_n=\langle n\rangle^{-1}\psi(z)$,
\begin{equation}
S\;\simeq\;\ln\langle n\rangle-\int_0^{\infty}\!\psi(z)\ln\psi(z)\,dz.
\label{eq:S-cont}
\end{equation}
Writing $\psi(z)=e^{-z}[1+f_s(z)]$ and expanding $\ln\psi$ in powers of $f_s$ on the support range $\mathcal{S}=[z_{\min},z_{\max}]$ of the measured distribution, the linear-in-$f_s$ contribution is controlled by the normalisation conditions $\int_{\mathcal S}\psi\,dz=1$ and $\langle z\rangle=1$; in the continuum limit it cancels identically, and on the finite support range it combines with the $\langle z\rangle=1$ constant and is reabsorbed into the factor $I_0$ introduced below. What remains is a correction quadratic in $f_s$. The calculation is given in Appendix~\ref{app:entropy}; the result is the central formula of this work,
\begin{equation}
S\;=\;\ln\langle n\rangle+I_0-\frac{1}{2}\!\int_{\mathcal{S}}\!e^{-z}f_s^2(z)\,dz+\mathcal{O}(f_s^3),
\label{eq:S-formula}
\end{equation}
with $I_0\equiv\int_{\mathcal S}e^{-z}dz$, evaluated below as the corresponding sum over the measured bins (the same support range as $S_{\rm dir}$). In the idealised continuum limit $z_{\min}\to0$, $z_{\max}\to\infty$, the support factor $I_0\to1$ and Eq.~\eqref{eq:S-formula} reduces to
\begin{equation}
S\;=\;\ln\langle n\rangle+1-\frac{1}{2}\!\int_0^\infty\!e^{-z}f_s^2(z)\,dz+\mathcal{O}(f_s^3).
\label{eq:S-ideal}
\end{equation}

Equations~\eqref{eq:S-formula} and~\eqref{eq:S-ideal} make precise, and explain, the empirical robustness of $S\simeq\ln\langle n\rangle+1$. The leading term $\ln\langle n\rangle+I_0$ is the geometric-cascade entropy: the Kharzeev--Levin value $\ln\langle n\rangle$~\cite{Kharzeev:2017} together with the entropy of the exponential KNO shape, which equals unity in the continuum limit. On the finite support range that constant is replaced by the support factor $I_0$ through the normalisation identities of Appendix~\ref{app:entropy}. The KNO-violating function $f_s$ enters only at second order: the term linear in $f_s$, which would have set the size of the deviation at $|f_s|\sim0.2$--$0.4$, cancels in the continuum limit and is otherwise reabsorbed into the leading term. The cancellation is a consequence of normalisation and $\langle z\rangle=1$ alone, and does not use the reciprocal symmetry $f_s(z)=f_s(1/z)$ studied in Secs.~\ref{sec:tower}--\ref{sec:test}; the entropy result is therefore independent of, and complementary to, the local constraints derived there. The leading deviation
\begin{equation}
\Delta S\;\equiv\;-\frac{1}{2}\!\int_{\mathcal{S}}\!e^{-z}f_s^2(z)\,dz
\label{eq:DeltaS}
\end{equation}
is negative-definite: any deviation of $\langle n\rangle P_n$ from the cascade exponential lowers the entropy below the geometric-cascade value $\ln\langle n\rangle+I_0$. It is the entropy deficit of the measured distribution relative to that value.

The deficit $\Delta S$ is dynamics-independent (within the KNO continuum) and data-driven: it is computed from the measured multiplicity shape $f_s$, with no assumption on the dynamics that generates it. It is the entropy gap between the measured $S$ and the geometric-cascade value $\ln\langle n\rangle+I_0$, a number that can be extracted from data and compared across energies and processes. The maximal-entanglement hypothesis can be examined in these terms: it predicts that $S$ lies below $\ln\langle n\rangle+I_0$ by the amount~\eqref{eq:DeltaS}, which can be set against the parton-level entropy of the BFKL~\cite{Hentschinski:2022,Hentschinski:2024} and dipole-cascade~\cite{Kutak:2025,Kutak:2025-2} calculations.

\begin{figure*}[t]
\includegraphics[width=\textwidth]{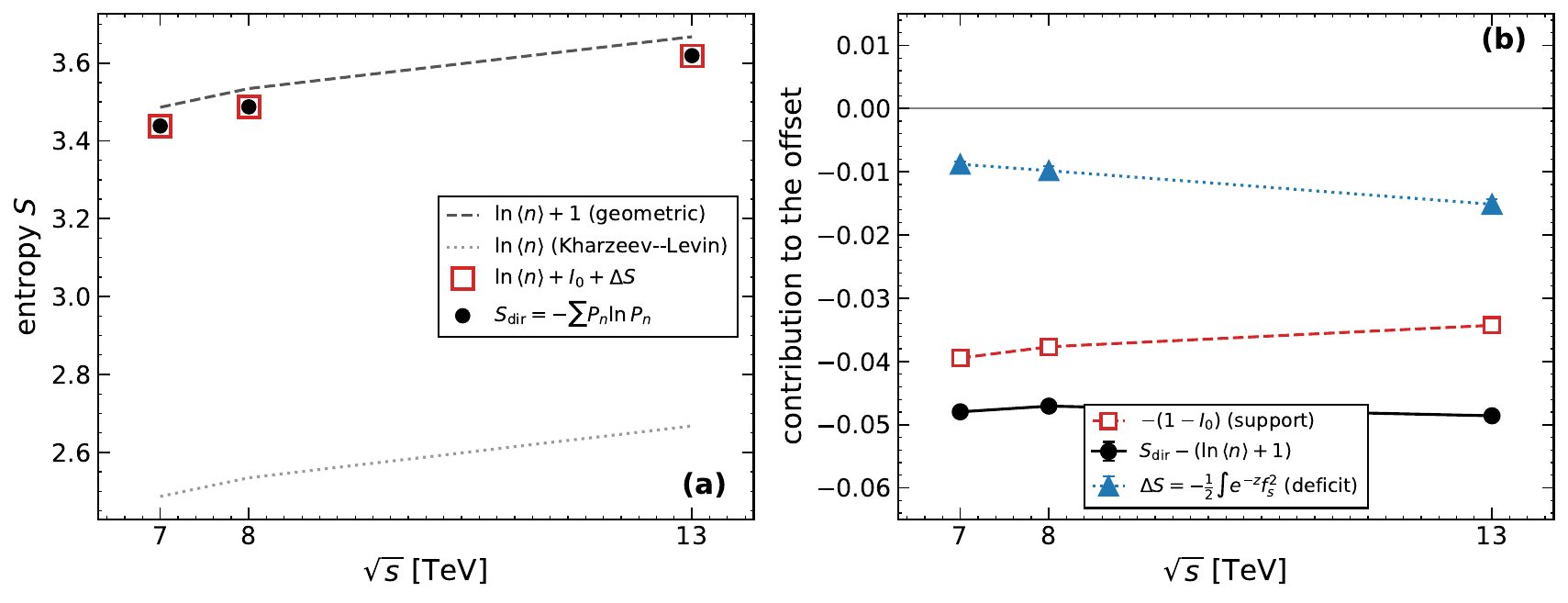}
\caption{(a) Shannon entropy of the charged-multiplicity distribution versus $\sqrt{s}$: the direct Shannon value $S_{\rm dir}=-\sum_n P_n\ln P_n$ (filled circles), the prediction $\ln\langle n\rangle+I_0+\Delta S$ of Eq.~\eqref{eq:S-formula} (open squares), the geometric-cascade value $\ln\langle n\rangle+1$ (dashed) and the Kharzeev--Levin parton-counting value $\ln\langle n\rangle$ (dotted)~\cite{Kharzeev:2017}. The data lie on $\ln\langle n\rangle+1$, well above $\ln\langle n\rangle$. (b) Decomposition of the offset $S_{\rm dir}-(\ln\langle n\rangle+1)$ (black circles) into the support correction $-(1-I_0)$ (red squares) and the quadratic deficit $\Delta S=-\tfrac12\int_{\mathcal S}e^{-z}f_s^2\,dz$ (blue triangles). Both pieces are of order $10^{-2}$, the support term being the larger by a factor $2$--$5$, so both must be kept in Eq.~\eqref{eq:S-formula}. Error bars on $\Delta S$ and on the offset are Monte Carlo bin-resampling uncertainties (statistical and systematic, treated as uncorrelated; a lower bound).}
\label{fig:entropy-result}
\end{figure*}

\subsection{Comparison with data}

We evaluate Eq.~\eqref{eq:S-formula} on the ATLAS charged-multiplicity data at $\sqrt{s}=7$, $8$ and $13$~TeV. The quadratic deficit is small,
\begin{equation}
\Delta S=-0.0088(5),\;-0.0098(7),\;-0.0151(8) \quad (7,\,8,\,13~{\rm TeV}),
\end{equation}
more than an order of magnitude smaller than the naive linear estimate $\sim|f_s|$, confirming that the cancellation of the linear term is what protects $S\simeq\ln\langle n\rangle+1$. The support factor is $I_0\simeq0.96$--$0.97$; the offset $1-I_0\simeq0.04$ is the leading Euler--Maclaurin correction to the $\sum_n\to\int dn$ step, set by the finite lower edge $z_{\min}=1/(2\langle n\rangle)$ of the lowest ($n=1$) bin, so that $-(1-I_0)\simeq-1/(2\langle n\rangle)$. The full offset $S_{\rm dir}-(\ln\langle n\rangle+1)=-(1-I_0)+\Delta S$ also carries the quadratic deficit; for $\langle n\rangle\simeq12$--$14$ both pieces are of the same order, the support term being the larger by a factor $2$--$5$, and must be kept for Eq.~\eqref{eq:S-formula} to be quantitatively meaningful. The uncertainties on $\Delta S$ follow from a Monte Carlo resampling of the bin contents within their combined statistical and systematic errors, treated as bin-to-bin uncorrelated; since the published systematics are correlated across bins (Appendix~\ref{app:fit}), they are a lower bound. The deficit grows with energy, the $13$~TeV value exceeding the $7$ and $8$~TeV ones by several standard deviations, while the latter two agree within their uncertainties.

\begin{figure}[!htbp]
\includegraphics[width=\columnwidth]{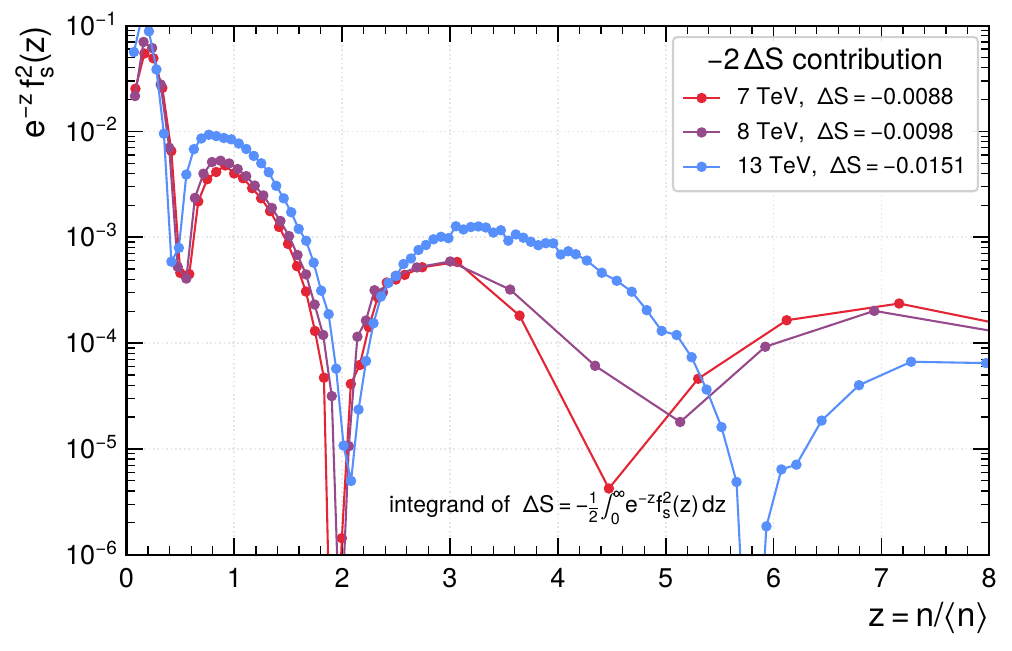}
\caption{The integrand $e^{-z}f_s^2(z)$ of the entropy deficit~\eqref{eq:DeltaS} for ATLAS data at $\sqrt{s}=7$, $8$, $13$~TeV. The full integral $\Delta S=-\tfrac12\int_{\mathcal S}e^{-z}f_s^2\,dz$ for each energy is given in the legend. Between $68\%$ and $73\%$ of $|\Delta S|$ comes from $z<1/3$ (the lowest multiplicity bins, where $\langle n\rangle P_n$ overshoots $e^{-z}$ and $f_s$ is large and positive), and $\sim90\%$ from $z<1.5$; the far tail $z>6$ contributes $\lesssim5\%$ ($4.1\%$, $2.5\%$, $0.7\%$ at $7$, $8$, $13$~TeV). The logarithmic $y$ axis enhances the small high-$z$ contributions relative to the dominant low-$z$ peak; the sharp downward spikes are the zeros of $f_s$, where the integrand $e^{-z}f_s^2$ vanishes.}
\label{fig:entropy}
\end{figure}

The prediction $\ln\langle n\rangle+I_0+\Delta S$ reproduces the direct Shannon entropy $S_{\rm dir}=-\sum_n P_n\ln P_n$ to $10^{-3}$ at all three energies (Table~\ref{tab:entropy} and Fig.~\ref{fig:entropy-result}), all three lying on the geometric-cascade line $\ln\langle n\rangle+1$ and well above the Kharzeev--Levin value $\ln\langle n\rangle$. Since both estimators use the same binned support range, this agreement tests the truncation of the $f_s$-expansion at second order (the residual between the prediction and $S_{\rm dir}$ is below $10^{-3}$ at all three energies), not the continuum step. The integrand of $\Delta S$ is concentrated at small $z$: between $68\%$ and $73\%$ of $|\Delta S|$ comes from $z<1/3$, where the lowest-multiplicity bins make $\langle n\rangle P_n$ exceed $e^{-z}$ and $f_s$ is large and positive, and roughly $90\%$ from $z<1.5$. The high-$z$ tail ($z\gtrsim6$), where $f_s\to-1$, is suppressed by the $e^{-z}$ weight and contributes $\lesssim5\%$ (Fig.~\ref{fig:entropy}).

The conclusion is a quantitative one. The geometric-cascade value $\ln\langle n\rangle+1$ is recovered from the data not because the multiplicity distribution is exponential, but because the leading KNO-violating correction to the entropy cancels by normalisation, leaving a deficit of only $|\Delta S|\lesssim0.015$. The deficit is itself a measurable number, computed from the multiplicity shape alone, and can be compared with the parton-level entanglement entropy of the BFKL and dipole-cascade approaches~\cite{Hentschinski:2022,Hentschinski:2024,Kutak:2025,Kutak:2025-2}.

\begin{table}[!htbp]
\caption{Entanglement entropy from the leading exponential, the support factor $I_0=\int_{\mathcal{S}}e^{-z}dz$ (evaluated as the bin sum $\sum_i e^{-n_i/\langle n\rangle}\,\Delta n_i/\langle n\rangle$ over the measured support, as for $S_{\rm dir}$ and $\Delta S$, so that it differs from the continuum $e^{-z_{\min}}$ at the $10^{-3}$ level), the quadratic deficit $\Delta S=-\tfrac{1}{2}\int_{\mathcal{S}}e^{-z}f_s^2 dz$, and the direct evaluation $S_{\rm dir}=-\sum_n P_n\ln P_n$ on the ATLAS data. The prediction $\ln\langle n\rangle+I_0+\Delta S$ agrees with $S_{\rm dir}$ at the $10^{-3}$ level. The direct entropy $S_{\rm dir}$ is evaluated at integer multiplicity: unit-width bins contribute $-P_n\ln P_n$ directly, and the wide high-$n$ bins are taken at flat per-unit density (the rectangular rule), so that $S_{\rm dir}$ is the integer-level Shannon entropy of the multiplicity distribution. The parenthetical uncertainty on $\Delta S$ (last digit) is from a Monte Carlo resampling of the bins within their statistical and systematic errors, treated as uncorrelated, and is a lower bound.}
\label{tab:entropy}
\begin{ruledtabular}
\begin{tabular}{c c c c c c}
$\sqrt{s}$ & $\langle n\rangle$ & $I_0$ & $\Delta S$ & $\ln\!\langle n\rangle\!+\!I_0\!+\!\Delta S$ & $S_{\rm dir}$ \\
\hline
$7$ TeV  & $12.02$ & $0.9605$ & $-0.0088(5)$ & $3.438$ & $3.438$\\
$8$ TeV  & $12.61$ & $0.9623$ & $-0.0098(7)$ & $3.487$ & $3.487$\\
$13$ TeV & $14.40$ & $0.9657$ & $-0.0151(8)$ & $3.618$ & $3.619$\\
\end{tabular}
\end{ruledtabular}
\end{table}

\section{Conclusion}
\label{sec:concl}

The reciprocal symmetry $f_s(z)=f_s(1/z)$ found in~\cite{Ouchen:2026} is studied in the present paper along three lines: a tower of local algebraic constraints on $P(n)$, a test against multiplicative-noise extensions of the cascade of the Mueller colour-dipole model, and a dynamics-independent expression (within the KNO continuum) for the entanglement entropy.

The substitution $h(u)=f_s(e^u)$ with $u=\ln z$ turns the symmetry into an evenness condition on $h$ and gives an infinite tower of local algebraic constraints on $P(n)$ at $n=\langle n\rangle$. The first member ($k=0$) reproduces the relation $P'(\langle n\rangle)=-P(\langle n\rangle)/\langle n\rangle$ of~\cite{Ouchen:2026}, while the second ($k=1$) reads $\langle n\rangle^3 P'''(\langle n\rangle)+6\langle n\rangle^2 P''(\langle n\rangle)=5P(\langle n\rangle)$ once the $k=0$ relation is imposed, the unconditional test being the residual $\delta_3=1-2\rho_0+\rho_1=0$. The $k=0$ test is passed by any $f_s$ stationary at $z=1$ (the condition $f_s'(1)=0$, i.e.\ $e^{z}\langle n\rangle P_n$ extremal at the mean), so it is a weak indicator of the symmetry, whereas $\delta_3=0$ is an independent constraint. In the ATLAS data the residual $\delta_3$ comes with a sizeable systematic uncertainty: at $13$~TeV it crosses zero monotonically as the fit window is enlarged (from $+0.79$ at $W=4$ to $-0.30$ at $W=8$), and the lower energies do not improve the situation. The global $\chi^2$ test is consistent with the symmetry at $7$ and $8$~TeV. At $13$~TeV the value is large, and a closure test (Appendix~\ref{app:fit}) shows it is not a binning artefact, so the data carry a residual pointwise deviation from the symmetry at the highest energy; whether this is a genuine breaking or a correlated detector systematic, not published differentially, is left open.

A multiplicative log-symmetric noise on the cascade rate of the Mueller colour-dipole model produces $f_s\propto z^2-4z+2$, which is not invariant under $z\to 1/z$. The same negative result holds for two-component geometric mixtures and for the negative binomial distribution. The dynamical origin of the reciprocal symmetry thus lies beyond simple extensions of the geometric cascade, and the tower of local constraints provides a quantitative filter for candidate models.

One of the main results of the present paper is the entanglement entropy formula $S=\ln\langle n\rangle+I_0-\tfrac{1}{2}\int_{\mathcal{S}}e^{-z}f_s^2 dz+\mathcal{O}(f_s^3)$, with $I_0=\int_{\mathcal{S}}e^{-z}dz$ and $\mathcal{S}$ the support of the measured distribution. It holds at leading order in the KNO-violating term and follows from normalisation and $\langle z\rangle=1$, independent of the symmetry. In the continuum limit $\mathcal{S}\to(0,\infty)$ one has $I_0\to 1$ and the simpler form is recovered. The numerical evaluation on ATLAS data agrees with the direct entropy at the $10^{-3}$ level.

Several questions remain open. A cascade-level origin of the reciprocal symmetry remains to be identified: the recombination and conformal-weight cascades of~\cite{Kutak:2025,Kutak:2025-2} reduce to a negative binomial that breaks it, and the AGK extension~\cite{Ouchen:2025AGK} carries it only at the reciprocal points $z=1/2,2$, so the dynamics that realises it in full must lie beyond these. The link between the discrete $\mathbb{Z}_2$ acting on $z$ and the M\"obius ($SL(2,\mathbb{C})$) invariance of the BFKL kernel is an interesting question on its own. Tests of $\rho_k$ in DIS~\cite{H1:2021}, $e^+e^-$, $pA$ and $AA$ data would clarify whether the symmetry is universal or specific to high-energy $pp$ collisions.

\begin{acknowledgments}
We thank Sergey Bondarenko for inspiring discussions. This work is supported in part by the ``Program of HEP support -- Council of Higher Education of Israel.''
\end{acknowledgments}

\appendix

\section{Derivation of the tower of local constraints}
\label{app:tower}

The reciprocal symmetry $f_s(z)=f_s(1/z)$ is equivalent to the statement that
\begin{equation}
h(u) \;\equiv\; f_s(e^u)
\end{equation}
is an even function of $u$. All odd-order derivatives of an even function vanish at the origin:
\begin{equation}
h^{(2k+1)}(0)\;=\;0,\qquad k=0,1,2,\ldots.
\label{eq:tower}
\end{equation}
Each of these conditions is a separate local constraint on the multiplicity distribution at $n=\langle n\rangle$.

The reciprocal symmetry is a discrete ($\mathbb{Z}_2$) statement and not a continuous one. Under the involution $u\to -u$ on the variable $u=\ln z$, any function $h(u)$ splits into even and odd parts, $h=h_{\rm even}+h_{\rm odd}$, and the symmetry sets $h_{\rm odd}\equiv 0$. This fact is the origin of the tower \eqref{eq:tower}. Namely, each odd derivative of $h$ at the fixed point $u=0$ projects onto $h_{\rm odd}$ and must therefore vanish. Being a discrete symmetry, it generates selection rules among the local derivatives of $P$ rather than a Noether current.

\subsection{Stirling-number expansion}

The chain rule gives $d/du=z\,d/dz$. The standard identity
\begin{equation}
\left(z\frac{d}{dz}\right)^n \;=\; \sum_{k=1}^{n} S(n,k)\,z^k\frac{d^k}{dz^k}
\end{equation}
where $S(n,k)$ are Stirling numbers of the second kind, yields
\begin{equation}
h^{(n)}(0)\;=\;\sum_{k=1}^{n} S(n,k)\,f_s^{(k)}(1).
\label{eq:hn-stirling}
\end{equation}
With $g(z)\equiv e^{z}\langle n\rangle P(z\langle n\rangle)$ so that $f_s=g-1$, one has $f_s^{(k)}(1)=g^{(k)}(1)$ for $k\geq 1$. The Leibniz rule applied to $g(z)=e^{z}q(z)$ with $q(z)=\langle n\rangle P(z\langle n\rangle)$ and $q^{(j)}(1)=\langle n\rangle^{j+1}P^{(j)}(\langle n\rangle)$ gives
\begin{equation}
g^{(k)}(1)\;=\;e\,\langle n\rangle\sum_{j=0}^{k}\binom{k}{j}\langle n\rangle^{j}P^{(j)}(\langle n\rangle).
\label{eq:gk-leibniz}
\end{equation}
We define the dimensionless quantities
\begin{equation}
\xi_j \;\equiv\; \frac{\langle n\rangle^{j}P^{(j)}(\langle n\rangle)}{P(\langle n\rangle)},\qquad \xi_0=1,
\end{equation}
and 
\begin{equation}
A_k\;\equiv\;\sum_{j=0}^{k}\binom{k}{j}\xi_j\;=\;\frac{g^{(k)}(1)}{e\,\langle n\rangle\,P(\langle n\rangle)}.
\end{equation}
The constraint \eqref{eq:tower} for odd $n=2k+1$ then takes the form
\begin{equation}
\sum_{m=1}^{2k+1} S(2k+1,m)\,A_m\;=\;0.
\label{eq:tower-A}
\end{equation}

\subsection{Explicit constraints}

We work out the first three members of the tower in turn: the $\boldsymbol{k=0}$ constraint (which recovers the previously established result), the new $\boldsymbol{k=1}$ constraint, and the $\boldsymbol{k=2}$ constraint, the last of which exposes the general pattern.

\paragraph*{$\boldsymbol{k=0}$.} Since $S(1,1)=1$,
\begin{equation}
A_1=0\;\Longleftrightarrow\;1+\xi_1=0\;\Longleftrightarrow\;P'(\langle n\rangle)=-\frac{P(\langle n\rangle)}{\langle n\rangle}.
\label{eq:k0}
\end{equation}
This recovers the result of~\cite{Ouchen:2026}.

\paragraph*{$\boldsymbol{k=1}$.} With $S(3,1)=1$, $S(3,2)=3$, $S(3,3)=1$ and $A_1=0$ already imposed,
\begin{equation}
A_3+3A_2=0.
\end{equation}
Substituting the explicit forms,
\begin{equation}
4+9\xi_1+6\xi_2+\xi_3\;=\;0.
\end{equation}
Here $A_1=0$ has already been used. The fully unconditional form of $h^{(3)}(0)=0$, before imposing the $k=0$ constraint, reads $5+10\xi_1+6\xi_2+\xi_3=0$, which is $5\delta_3=0$ in terms of the residual of Eq.~\eqref{eq:delta3}. The two differ by $\xi_1+1$, which vanishes on the $k=0$ relation $\xi_1=-1$, so that both reduce to Eq.~\eqref{eq:k1-xi}. Using $\xi_1=-1$, this reduces to
\begin{equation}
\xi_3+6\,\xi_2\;=\;5
\label{eq:k1-xi}
\end{equation}
or, in terms of $P$, this is the relation~\eqref{eq:k1} quoted in the main text.

\paragraph*{$\boldsymbol{k=2}$.} The $S(5,m)$ coefficients are $(1,15,25,10,1)$. After using the lower-order constraints, one obtains
\begin{equation}
\xi_5+15\,\xi_4\;=\;290\,\xi_2-276,
\label{eq:k2-xi}
\end{equation}
which involves $\xi_2$, $\xi_4$ and $\xi_5$ together (obtained from the unconditional $n=5$ constraint after using $\xi_1=-1$ and $\xi_3+6\xi_2=5$). The $k=2$ relation is therefore not closed in $\xi_4,\xi_5$ alone but ties them to $\xi_2$ via the $k=1$ relation. The general structure of the tower is that the unconditional $k$-th constraint mixes $\xi_1,\ldots,\xi_{2k+1}$ with Stirling-number coefficients; once the lower-order relations are imposed it reduces to a combination of fewer ratios (for $k=2$, of $\xi_2,\xi_4,\xi_5$).

\subsection{Methodological remark}

The Gaussian-in-$\ln z$ parametrisation of \cite{Ouchen:2026},
\begin{equation}
f_s^{\rm fit}(z) = a + b\,e^{-c(\ln z-\mu)^2},
\label{eq:gauss-fit}
\end{equation}
satisfies $h_{\rm fit}(u)=h_{\rm fit}(-u)$ for $\mu=0$ identically, and therefore satisfies all $\rho_k=1$ by construction whenever $\mu=0$. A non-trivial test of $\rho_1$ requires a parametrisation that does not impose the symmetry. A local polynomial fit in the variable $n$,
\begin{equation}
P^{\rm fit}(n)=\sum_{k=0}^{N} c_k\,\frac{(n-\langle n\rangle)^k}{k!},\qquad c_k\simeq P^{(k)}(\langle n\rangle),
\label{eq:poly-fit}
\end{equation}
provides such an extraction that does not impose the reciprocal symmetry by construction. We use this approach in Sec.~\ref{sec:test}, with the fit details collected in Appendix~\ref{app:fit}. In this language the evenness of $h(u)=f_s(e^u)$ is the statement that the KNO-violating term is symmetric in $\ln z$, i.e.\ a near-log-normal deformation of the geometric KNO function. It is worth mentioning that lognormal parametrisations of multiplicity distributions have a long history~\cite{CarruthersShih:1983,DreminGary:2001}.

\section{Local-fit extraction of the third-derivative residual}
\label{app:fit}

The $\delta_3$ extraction summarised in Sec.~\ref{sec:test} is described here in detail.

\subsection{Fit method}

Because $P_n$ is a discrete distribution and $\langle n\rangle$ is non-integer, the derivatives $P^{(k)}(\langle n\rangle)$ are not defined pointwise. We interpret them as derivatives of a smooth local interpolation of the binned distribution near $n=\langle n\rangle$. For each energy we fit the unit-width bins lying within $|n-\langle n\rangle|\le W$ with the truncated Taylor polynomial
\begin{equation}
P^{\rm fit}(n)=\sum_{k=0}^{N} c_k\,\frac{(n-\langle n\rangle)^k}{k!},
\qquad c_k\simeq P^{(k)}(\langle n\rangle),
\label{eq:poly-fit-app}
\end{equation}
by weighted linear least squares, with weights $1/\sigma_i^2$ and $\sigma_i^2=\sigma_{{\rm stat},i}^2+\sigma_{{\rm sys},i}^2$ the in-quadrature sum of the statistical and systematic bin uncertainties. The mean $\langle n\rangle$ is computed directly from the same binned distribution, $\langle n\rangle=\sum_n n\,P_n\,w_n$, giving $12.02$, $12.61$ and $14.40$ at $7$, $8$ and $13$~TeV, in agreement with~\cite{Kulchitsky:2022}. Here $n$ is taken at the integer bin label, with the wide high-$n$ bins entering at flat per-unit density; bin centres would shift $\langle n\rangle$ by $\sim0.3\%$. These binning-consistent values guarantee $\int z\,\psi(z)\,dz=1$ within the analysis. This fact underlies the linear-term cancellation in the entropy formula of Sec.~\ref{sec:entropy}. This parametrisation does not impose the reciprocal symmetry, unlike the Gaussian-in-$\ln z$ form of~\cite{Ouchen:2026}, which satisfies all $\rho_k=1$ by construction whenever its centre is at $\mu=0$.

The fit coefficients give
\begin{equation}
\rho_0=-\langle n\rangle\frac{c_1}{c_0},\qquad
\rho_1=\frac{\langle n\rangle^3 c_3+6\langle n\rangle^2 c_2}{5\,c_0},
\end{equation}
and $\delta_3=1-2\rho_0+\rho_1$. The within-fit uncertainty on $\delta_3$ is obtained by propagating the full covariance matrix of the $c_k$ (equivalently, by a Monte Carlo resampling of the bin contents within their errors); both give identical results to the quoted precision. These errors treat the bins as independent and are therefore a lower bound on the true uncertainty, since the published systematic uncertainties are correlated across bins, as discussed below.

\subsection{Window and degree dependence}

The third derivative entering $\rho_1$ is poorly conditioned: it is read off a degree-$N$ polynomial fit to $\sim2W+1$ unit bins, and is sensitive to both $W$ and $N$. Table~\ref{tab:delta3-window} gives $\delta_3$ versus $W$ at fixed degree $N=4$; Table~\ref{tab:delta3-degree} gives $\delta_3$ versus $N$ at fixed $W=6$, both for the $13$~TeV data where the experimental errors are smallest. At $13$~TeV $\delta_3$ falls monotonically through zero as $W$ grows, while the degree sweep does not converge. The $\chi^2/\mathrm{dof}$ of the fits ($N=3$: $2.1$, $N=4$: $1.0$, $N=5$: $1.1$, $N=6$: $0.3$ at $W=6$) favours $N=4$ over $N=3,5$ and indicates that $N=6$ overfits the data, while $N=4$ is not robust against the statistically comparable $N=5$. The resulting modelling spread of $\delta_3$ is of order $\pm0.5$, several times the within-fit error, and dominates the uncertainty on the third-derivative test. The full $(W\times N)$ grid for all three energies is given in the supplementary file \texttt{delta3\_grid.csv}.

\begin{figure*}[t]
\includegraphics[width=\textwidth]{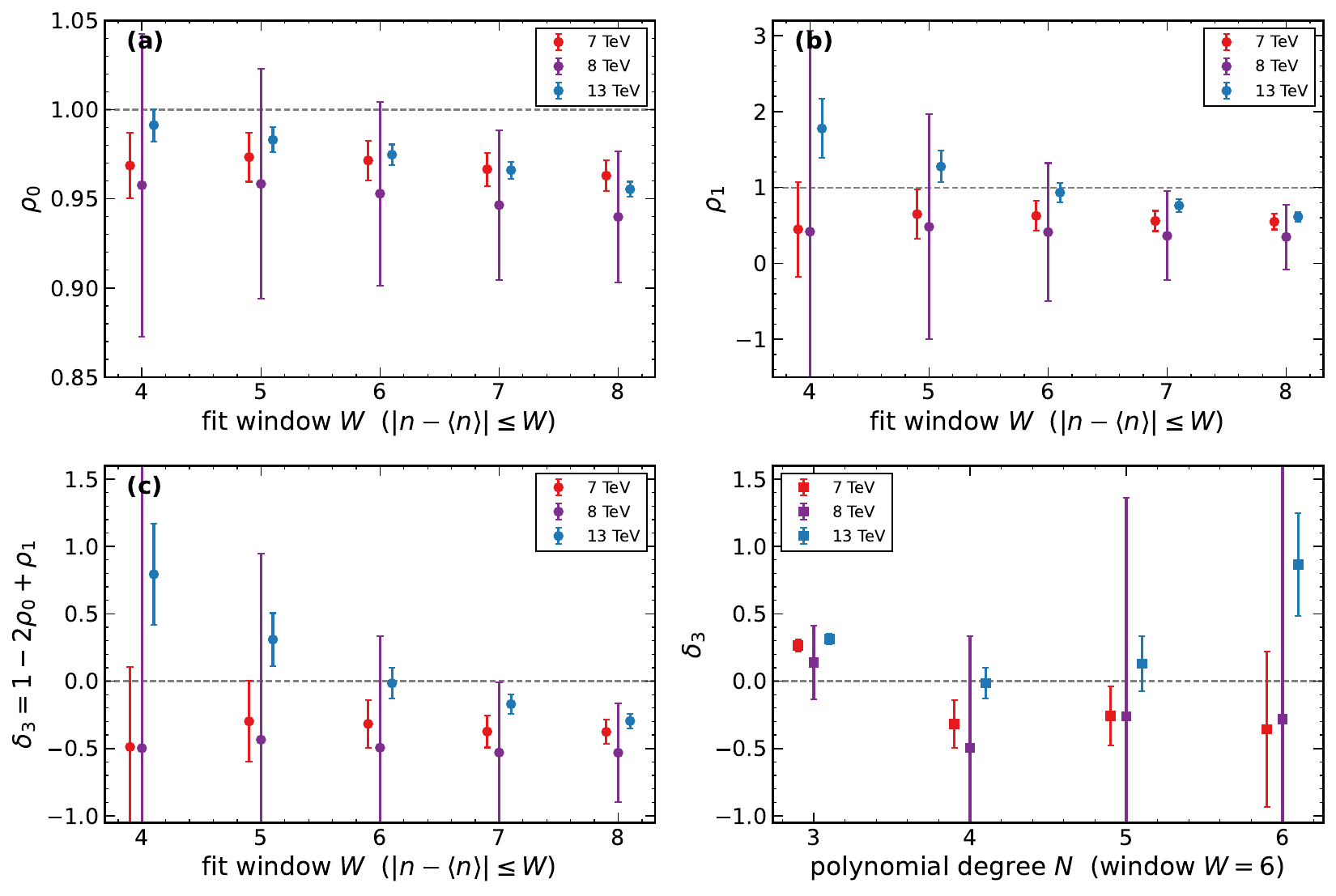}
\caption{Robustness of the local-fit ratios for ATLAS data at $\sqrt{s}=7$, $8$, $13$ TeV. (a) $\rho_0$ and (b) $\rho_1$ as functions of the fit window $W$ at fixed degree $N=4$; (c) the residual $\delta_3=1-2\rho_0+\rho_1$ versus $W$; (d) $\delta_3$ versus polynomial degree $N$ at fixed $W=6$. The dashed lines mark the symmetric values $\rho_0=\rho_1=1$ and $\delta_3=0$. The ratio $\rho_0$ is stable across windows, whereas $\rho_1$, and hence $\delta_3$, vary strongly: at $13$~TeV $\delta_3$ falls monotonically through zero with $W$, and the degree sweep does not converge. Error bars are within-fit only and exclude the window/degree modelling spread. Markers are offset horizontally for readability.}
\label{fig:robustness}
\end{figure*}

Figure~\ref{fig:robustness} collects the window and degree dependence of $\rho_0$, $\rho_1$ and the residual $\delta_3$: $\rho_0$ is stable across windows, while $\rho_1$, set by the third derivative, varies strongly with the fit range, and $\delta_3$ inherits this instability, falling monotonically through zero at $13$~TeV.


\begin{table*}[t]
\caption{Window dependence of the third-derivative residual $\delta_3\equiv 1-2\rho_0+\rho_1$ from local degree-4 polynomial fits to the ATLAS data, for fit windows $|n-\langle n\rangle|\le W$. Errors are the within-fit (bin-uncorrelated) propagation only. At $13$~TeV $\delta_3$ decreases monotonically through zero as $W$ grows: the apparent ``consistency with zero'' near $W=6$ is a zero-crossing of this trend, not a stable plateau. The lower energies remain negative.}
\label{tab:delta3-window}
\begin{ruledtabular}
\begin{tabular}{c c c c c c}
$\sqrt{s}$ & $W=4$ & $W=5$ & $W=6$ & $W=7$ & $W=8$ \\
\hline
$13$ TeV & $+0.79\pm0.38$ & $+0.31\pm0.20$ & $-0.02\pm0.11$ & $-0.17\pm0.07$ & $-0.30\pm0.05$ \\
$8$ TeV & $-0.50\pm2.54$ & $-0.44\pm1.38$ & $-0.49\pm0.82$ & $-0.53\pm0.52$ & $-0.53\pm0.36$ \\
$7$ TeV & $-0.49\pm0.60$ & $-0.30\pm0.30$ & $-0.32\pm0.18$ & $-0.37\pm0.12$ & $-0.38\pm0.09$ \\
\end{tabular}
\end{ruledtabular}
\end{table*}

\begin{table}[t]
\caption{Polynomial-degree dependence of $\delta_3$ at $\sqrt{s}=13$~TeV for the fixed window $W=6$. The central value does not converge with $N$; the $\chi^2/\mathrm{dof}$ values ($N=3\!:\!2.1$, $N=4\!:\!1.0$, $N=5\!:\!1.1$, $N=6\!:\!0.3$) favour $N=4$ over $N=3,5$ but $N=6$ overfits. The $N=4$ result is therefore not robust against the statistically comparable $N=5$.}
\label{tab:delta3-degree}
\begin{ruledtabular}
\begin{tabular}{c c c c c}
 & $N=3$ & $N=4$ & $N=5$ & $N=6$ \\
\hline
$\delta_3$ & $+0.31\pm0.04$ & $-0.02\pm0.11$ & $+0.13\pm0.21$ & $+0.87\pm0.38$ \\
\end{tabular}
\end{ruledtabular}
\end{table}

\subsection{Validation on a closed-form distribution}

To check that the local-fit estimator is unbiased we apply it to a synthetic negative binomial distribution of known parameters $(k,\langle n\rangle)$, binned and cut ($n\ge1$) exactly as the ATLAS data. The negative binomial has a known, generally non-zero, true third-derivative residual $\delta_3^{\rm NBD}(k,\langle n\rangle)$ computed analytically from the digamma/polygamma derivatives of the Gamma-function continuation of $P_n$. For $\langle n\rangle\simeq14.4$ and $k=2,3,5$ the estimator recovers $\delta_3^{\rm NBD}$ with a bias $\lesssim0.07$ for $k=2,3$ and growing to $\sim0.2$ at $k=5$ over the windows $W=4$--$6$ used here (Table~\ref{tab:nbd-recovery}), confirming that the estimator is approximately unbiased and that the strong window dependence seen in the ATLAS data at $13$~TeV reflects the data, not a pathology of the estimator.

\begin{table}[t]
\caption{Validation of the local-fit estimator on a synthetic negative binomial distribution of mean $\langle n\rangle=14.4$, binned and cut ($n\ge1$) as the ATLAS data. $\delta_3^{\rm true}$ is computed analytically from the digamma/polygamma derivatives of the Gamma-function continuation of $P_n$ at the effective mean; $\delta_3^{\rm fit}$ is the degree-$4$ local-fit estimate. The estimator recovers the true (non-zero) residual with a bias $\lesssim0.07$ for $k=2,3$ and $\sim0.2$ for $k=5$ over the windows $W=4$--$6$ used in the analysis. The negative binomial satisfies the $k=1$ constraint ($\delta_3=0$) only in the degenerate geometric limit.}
\label{tab:nbd-recovery}
\begin{ruledtabular}
\begin{tabular}{c c c c c}
$k$ & $\delta_3^{\rm true}$ & $\delta_3^{\rm fit}(W{=}4)$ & $W{=}5$ & $W{=}6$ \\
\hline
$2$ & $-0.208$ & $-0.180$ & $-0.163$ & $-0.142$ \\
$3$ & $-0.445$ & $-0.434$ & $-0.430$ & $-0.428$ \\
$5$ & $-0.938$ & $-1.017$ & $-1.073$ & $-1.146$ \\
\end{tabular}
\end{ruledtabular}
\end{table}

\subsection{Global symmetry test}

A quantitative measure of the global symmetry is given by
\begin{equation}
\chi^2_{\rm sym}\;\equiv\;\sum_i\frac{[f_s(z_i)-f_s(1/z_i)]^2}{\sigma^2(z_i)+\sigma^2(1/z_i)},
\qquad \sigma^2\equiv\sigma_{\rm stat}^2+\sigma_{\rm sys}^2,
\label{eq:chi2-sym}
\end{equation}
summing over data points $z_i$ with $1/3<z_i<3$ (each pair counted once), where $f_s(1/z_i)$ is obtained by linear interpolation in $z$ and $\sigma(1/z_i)$ is propagated accordingly. We take as representatives the measured points with $z_i\ge1$ in the window, each paired once with its reciprocal, and obtain $\chi^2_{\rm sym}/N_{\rm pairs}=1.5$ at $7$~TeV ($N_{\rm pairs}=20$), $0.4$ at $8$~TeV ($N_{\rm pairs}=20$), and $27$ at $13$~TeV ($N_{\rm pairs}=29$); the recipe is reproducible from the supplementary script. The reciprocal symmetry is consistent with the data at $7$ and $8$~TeV. The large value at $13$~TeV is examined by the closure test below.

\begin{figure}[t]
\includegraphics[width=\columnwidth]{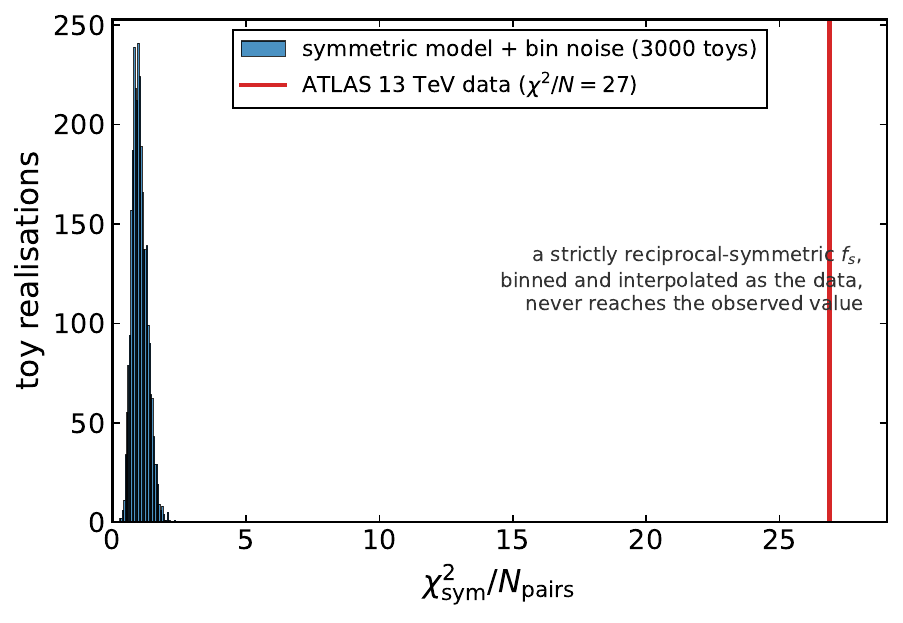}
\caption{Closure test of the global symmetry estimator at $\sqrt{s}=13$~TeV. A strictly reciprocal-symmetric $f_s$ (even in $u=\ln z$), sampled on the measured $z$ grid, perturbed within the published bin errors, and passed through the same interpolation and pairing as the data, gives $\chi^2_{\rm sym}/N_{\rm pairs}$ clustered near unity (histogram, $3000$ realisations). The measured value (red line) lies far outside this distribution, so the large $13$~TeV $\chi^2_{\rm sym}$ is not an artefact of the binning or the interpolation.}
\label{fig:closure}
\end{figure}

To test whether the large $13$~TeV value is an artefact of the binning and the interpolation, we apply the same estimator to a strictly reciprocal-symmetric reference. A smooth even function $h(u)=h(-u)$, a Gaussian in $u=\ln z$ fitted to the $13$~TeV $f_s$, is evaluated on the measured $z$ grid, perturbed bin-by-bin within the published errors, and passed through the same interpolation and pairing. Over $3000$ realisations the symmetric reference gives $\chi^2_{\rm sym}/N_{\rm pairs}=1.0$ on average. Of these, $95\%$ fall below $1.5$ and none exceeds $2.4$ (Fig.~\ref{fig:closure}). The noiseless symmetric function gives $0.03$, so the interpolation by itself contributes negligibly. This ensemble is generated on the $13$~TeV grid and errors, and its percentiles do not carry over to the lower energies, whose grids and uncertainties differ. The measured value $27$ is therefore not reproduced by a symmetric distribution binned and interpolated as the data, and reflects a genuine pointwise deviation from $f_s(z)=f_s(1/z)$ at $13$~TeV. The $k=0$ ratio $\rho_0=0.975$ at $13$~TeV stays close to unity, so the deviation lies in the higher-order, pointwise structure of $f_s$ and not in its leading behaviour.

The errors in Eq.~\eqref{eq:chi2-sym} are the in-quadrature sum of statistical and systematic bin uncertainties, treated as bin-to-bin uncorrelated; the dominant systematic is in fact correlated across bins, but ATLAS does not publish it in differential form. A $z$-dependent correlated systematic can mimic such a deviation and cannot be excluded with the public data, so we do not claim a definitive breaking of the symmetry. The deviation is consistent with the unstable third-derivative residual $\delta_3$ at $13$~TeV (Sec.~\ref{sec:test}), and both indicate that at the highest energy the reciprocal symmetry is realised less precisely than at $7$ and $8$~TeV. It remains a good description of $f_s$ through the Gaussian-in-$\ln z$ form of Ref.~\cite{Ouchen:2026}; the present test probes it pointwise, at the level of the quoted bin errors.

\section{Derivation of the multiplicative-noise and related results}
\label{app:noise}

In the Mueller colour-dipole model~\cite{Mueller:1994,Mueller:1995} the multiplicity distribution is geometric,
\begin{equation}
P_n^{\rm Mueller}=e^{-\alpha Y}(1-e^{-\alpha Y})^{n-1},
\label{eq:Mueller}
\end{equation}
with mean $\langle n\rangle=e^{\alpha Y}$. In the KNO limit $\langle n\rangle\to\infty$ this gives $\langle n\rangle P_n\to e^{-z}$, hence $f_s\equiv 0$.

We consider an event-by-event fluctuation of the cascade rate, $\alpha\to\alpha(1+\sigma\nu)$ with $\nu$ a unit-variance random variable, $\langle\nu\rangle=0$ and (for definiteness in the leading-order expansion below) all odd moments $\langle\nu^{2k+1}\rangle=0$ for $k\geq 1$, as is the case for any symmetric noise distribution including a Gaussian. The event-mean is $\langle n\rangle_{\rm event}=\langle n\rangle_0\,e^{\eta}$ with $\langle n\rangle_0\equiv e^{\alpha Y}$ the baseline (noise-free) cascade mean and $\eta\equiv\sigma\nu\,\ln\langle n\rangle_0$ a noise variable with $\langle\eta\rangle=0$ and $\langle\eta^2\rangle\equiv\sigma_\eta^2$. The ensemble mean that enters the experimental KNO variable $z=n/\langle n\rangle$ is
\begin{equation}
\langle n\rangle\;=\;\langle\langle n\rangle_{\rm event}\rangle\;=\;\langle n\rangle_0\,\langle e^\eta\rangle\;=\;\langle n\rangle_0\!\left(1+\frac{\sigma_\eta^2}{2}\right)+\mathcal{O}(\sigma_\eta^4),
\label{eq:nbar-rescale}
\end{equation}
which differs from $\langle n\rangle_0$ at order $\sigma_\eta^2$ and must be tracked when rescaling the model prediction to experimental coordinates. Averaging the geometric distribution at fixed event,
\begin{equation}
\langle n\rangle_0\,P_n^{\rm event}\;=\;e^{-\eta}\exp(-z_0 e^{-\eta}),\qquad z_0\equiv n/\langle n\rangle_0,
\end{equation}
over the noise distribution and expanding to $\mathcal{O}(\sigma_\eta^2)$,
\begin{equation}
\langle\,\langle n\rangle_0 P_n\,\rangle\;=\;e^{-z_0}\!\left[1+\frac{\sigma_\eta^2}{2}(z_0^2-3z_0+1)\right]+\mathcal{O}(\sigma_\eta^4).
\label{eq:fixed-mean}
\end{equation}
Eq.~\eqref{eq:fixed-mean} is the answer in $z_0=n/\langle n\rangle_0$. To convert to the experimental variable $z=n/\langle n\rangle$, use \eqref{eq:nbar-rescale} so $z_0=z(1+\sigma_\eta^2/2)$, and multiply by the Jacobian $dz_0/dz=1+\sigma_\eta^2/2$ to preserve normalisation. Expanding to $\mathcal{O}(\sigma_\eta^2)$,
\begin{equation}
\langle n\rangle P_n\;=\;e^{-z}\!\left[1+\frac{\sigma_\eta^2}{2}(z^2-4z+2)\right]+\mathcal{O}(\sigma_\eta^4),
\end{equation}
so that $f_s=(\sigma_\eta^2/2)(z^2-4z+2)$ [Eq.~\eqref{eq:noise-fs}].
The polynomial $z^2-4z+2$ satisfies $\int_0^\infty e^{-z}(z^2-4z+2)\,dz=0$ and $\int_0^\infty z\,e^{-z}(z^2-4z+2)\,dz=0$, ensuring the normalisation $\int\psi dz=1$ and the constraint $\langle z\rangle=1$ that hold by construction in experimental data. Its roots, $z=2\pm\sqrt 2\approx 0.59,\,3.41$, are not reciprocals of each other (their product is $2$). It is therefore not invariant under $z\to 1/z$: the value at $z=2$ is $-2$, the value at $z=1/2$ is $1/4$. A multiplicative log-symmetric noise on the cascade rate does not reproduce $f_s(z)=f_s(1/z)$ at leading order.

The same conclusion holds for two-component geometric mixtures and for the negative binomial distribution. For a two-component mixture $P_n=w_1 G(n;\langle n_1\rangle)+w_2 G(n;\langle n_2\rangle)$ with $G(n;\bar m)$ the geometric distribution of mean $\bar m$, the resulting $f_s$ in the KNO limit takes the form $f_s(z)=-1+\sum_i w_i r_i e^{(1-r_i)z}$ with $r_i=\langle n\rangle/\langle n_i\rangle$, which is a sum of exponentials in $z$ and is not invariant under $z\to 1/z$ except in the degenerate case $r_1=r_2=1$, in which $f_s\equiv 0$. For the negative binomial of parameter $k$, the KNO-limit form $\langle n\rangle P_n = (k^k/\Gamma(k))z^{k-1}e^{-kz}$ gives $f_s(z)=-1+(k^k/\Gamma(k))z^{k-1}e^{-(k-1)z}$, which equals $f_s(1/z)$ only at $k=1$ (the geometric limit, $f_s\equiv 0$). For this form one finds $\rho_0=1$ identically and $\delta_3=(1-k)/5$, so that the negative binomial passes the $k=0$ relation for every $k$ while violating the $k=1$ constraint at any $k\neq 1$. The large-$k$ limit does not restore the symmetry. In this limit the KNO function narrows towards a $\delta$-distribution at $z=1$ and $f_s$ grows rather than vanishing.

The reciprocal symmetry observed in the data therefore constrains the cascade dynamics beyond rate fluctuations and beyond the standard mixture and negative-binomial extensions. The tower of local constraints \eqref{eq:tower-A} can be used as a filter for candidate cascade extensions: any model whose KNO-scaled multiplicity satisfies the reciprocal symmetry must, in particular, satisfy $\xi_3+6\xi_2=5$ at $n=\langle n\rangle$. The AGK extension of the chain of the Mueller colour-dipole model~\cite{Ouchen:2025AGK} illustrates the same point. At leading order in $1/\langle n\rangle$ the AGK distribution gives $\langle n\rangle P_n\to e^{-z}[1+2/\langle n\rangle]$, namely a \emph{constant} $f_s\simeq2/\langle n\rangle$. A constant $f_s$ is, however, incompatible with the identities $\int e^{-z}f_s\,dz=\int z\,e^{-z}f_s\,dz=0$ that follow from normalisation and $\langle z\rangle=1$ (and that we use in Secs.~\ref{sec:tower} and~\ref{sec:entropy}), since it would shift the normalisation by $2/\langle n\rangle$. This offset is precisely the bare-versus-true-mean difference that is reabsorbed by the rescaling to the physical variable $z=n/\langle n\rangle$ [cf.\ Eq.~\eqref{eq:nbar-rescale}], after which the leading constant $f_s$ vanishes. Its invariance under $z\to1/z$ is therefore empty, and the symmetry content is carried by the subleading $z$-dependence alone. At that order the AGK distribution does \emph{not} possess the full reciprocal symmetry. It has the reciprocal intersection points $z=1/2$ and $z=2$, where $f_s$ takes a common value, while invariance across the whole window is absent~\cite{Ouchen:2025AGK}. The role of $z=2$ as a fixed point of approximate KNO restoration in the ATLAS data is examined in~\cite{Ouchen:2025MEM}. The non-trivial $z$-dependence that carries the symmetry information thus requires dipole merging (Pomeron loops) or analogous higher-order contributions. Dipole evolution including recombination and transitions to the vacuum has been studied recently in~\cite{Kutak:2025,Kutak:2025-2}, and the local constraints~\eqref{eq:k1} answer the question for these schemes directly. In both the recombination cascade of~\cite{Kutak:2025} and the conformal-weight generalisation of~\cite{Kutak:2025-2} the pure Mueller ($k=1$) limit is geometric, $f_s\equiv0$, while a nonzero recombination rate or anomalous dimension narrows the distribution into a negative binomial whose KNO limit is the gamma form $\langle n\rangle P_n\to(k^k/\Gamma(k))\,z^{k-1}e^{-kz}$, with shape $k=2h$ set by the recombination rate or the conformal weight $h$. This is the negative-binomial case above: $\rho_0=1$ for all $k$, but $\delta_3=(1-k)/5$, so the reciprocal symmetry is broken for any $k\neq1$. At the value $k=2h\simeq1.84$ favoured by the multiplicity fit of~\cite{Kutak:2025-2} this gives $\delta_3\simeq-0.17$. The reciprocal symmetry is therefore a diagnostic of the geometric (Mueller) point, and recombination or a conformal-weight shift moves the cascade away from it; a scheme that preserves the symmetry must keep $\xi_3+6\xi_2=5$ at $n=\langle n\rangle$, which neither of these does. For the vacuum-transition branch of~\cite{Kutak:2025} the KNO scaling itself is lost, a finite weight accumulating at $n=0$ while $\langle n\rangle$ falls with rapidity, so the test applies to the surviving $n\ge1$ negative-binomial component.

\section{Entropy expansion and cancellation of the linear term}
\label{app:entropy}

This appendix derives Eqs.~\eqref{eq:S-formula} and~\eqref{eq:S-ideal} from the continuum entropy~\eqref{eq:S-cont}. We write the KNO-scaled distribution as
\begin{equation}
\psi(z)=e^{-z}\,\bigl[1+f_s(z)\bigr]
\end{equation}
on the support $\mathcal{S}=[z_{\min},z_{\max}]$ of $\psi$, with $z_{\min}$ the lower edge of the lowest measured bin ($z_{\min}=1/(2\langle n\rangle)$ for the $n=1$ bin) and $z_{\max}$ the upper edge of the highest. Expanding the logarithm,
\begin{equation}
\ln\psi=-z+\ln(1+f_s)=-z+f_s-\frac{f_s^2}{2}+\mathcal{O}(f_s^3),
\end{equation}
and multiplying by $\psi$,
\begin{align}
\psi\ln\psi
&= -z\,e^{-z}(1+f_s)+e^{-z}(1+f_s)\!\left(f_s-\frac{f_s^2}{2}+\cdots\right)\nonumber\\
&= -z\,e^{-z}(1+f_s)+e^{-z}\!\left(f_s+\frac{f_s^2}{2}+\mathcal{O}(f_s^3)\right).
\label{eq:psilnpsi}
\end{align}

The two normalisation properties of $\psi$, $\int_{\mathcal S}\psi\,dz=1$ and $\int_{\mathcal S}z\,\psi\,dz=1$, translate into integral identities for $f_s$,
\begin{align}
\int_{\mathcal{S}}\!\!\psi(z)\,dz &= 1\;\Longleftrightarrow\; \int_{\mathcal{S}}\!\!e^{-z}f_s(z)\,dz = 1-I_0,\\
\int_{\mathcal{S}}\!\!z\,\psi(z)\,dz &= 1\;\Longleftrightarrow\; \int_{\mathcal{S}}\!\!z\,e^{-z}f_s(z)\,dz = 1-I_1,
\end{align}
with $I_0\equiv\int_{\mathcal{S}}e^{-z}dz$ and $I_1\equiv\int_{\mathcal{S}}z\,e^{-z}dz$. In the idealised limit $z_{\min}\to0$, $z_{\max}\to\infty$ one has $I_0,I_1\to1$, and both identities reduce to the exact cancellation $\int e^{-z}f_s=\int z\,e^{-z}f_s=0$ used in Sec.~\ref{sec:tower}.

Integrating Eq.~\eqref{eq:psilnpsi} over $\mathcal{S}$ and inserting the identities, the constant term $-I_1$ from $-z\,e^{-z}$, the linear term $-(1-I_1)$ from $-z\,e^{-z}f_s$, and the linear term $+(1-I_0)$ from $e^{-z}f_s$ combine,
\begin{align}
\int_{\mathcal{S}}\psi\ln\psi\,dz
&= -I_1-(1-I_1)+(1-I_0)\nonumber\\
&\quad+\frac{1}{2}\!\int_{\mathcal{S}}e^{-z}f_s^2\,dz+\mathcal{O}(f_s^3)\nonumber\\
&= -I_0+\frac{1}{2}\!\int_{\mathcal{S}}e^{-z}f_s^2\,dz+\mathcal{O}(f_s^3).
\end{align}
The two linear-in-$f_s$ contributions cancel against the constant from $\langle z\rangle=1$, leaving $-I_0$; this is the cancellation that protects the entropy at first order. Substituting into Eq.~\eqref{eq:S-cont},
\begin{equation}
S\;=\;\ln\langle n\rangle+I_0-\frac{1}{2}\!\int_{\mathcal{S}}\!e^{-z}f_s^2(z)\,dz+\mathcal{O}(f_s^3),
\end{equation}
which is Eq.~\eqref{eq:S-formula}. In the idealised continuum limit $I_0\to\int_0^\infty e^{-z}dz=1$ and one recovers Eq.~\eqref{eq:S-ideal},
\begin{equation}
S\;=\;\ln\langle n\rangle+1-\frac{1}{2}\!\int_0^\infty\!e^{-z}f_s^2(z)\,dz+\mathcal{O}(f_s^3).
\end{equation}
The cancellation uses only normalisation and $\langle z\rangle=1$; it does not invoke the reciprocal symmetry $f_s(z)=f_s(1/z)$, so Eqs.~\eqref{eq:S-formula}--\eqref{eq:S-ideal} hold for any normalised KNO distribution with unit mean, independently of the constraints of Sec.~\ref{sec:tower}.

\end{document}